# A SILICON-BASED MICRO GAS TURBINE ENGINE FOR POWER GENERATION

X. C. Shan[1], Z. F. Wang[1], R. Maeda[2], Y. F. Sun[1] M. Wu[3] and J. S. Hua[3]

[1] Singapore Institute of Manufacturing Technology, Singapore 658075
[2] MEMS & Packaging Research Lab, AIST, Tsukuba, 305-8564 Japan
[3] Institute of High Performance Computing, Singapore 117528

## ABSTRACT

This paper reports on our research in developing a micro power generation system based on gas turbine engine and piezoelectric converter. The micro gas turbine engine consists of a micro combustor, a turbine and a centrifugal compressor. Comprehensive simulation has been implemented to optimal the component design. We have successfully demonstrated a silicon-based micro combustor, which consists of seven layers of silicon structures. A hairpin-shaped design is applied to the fuel/air recirculation channel. The micro combustor can sustain a stable combustion with an exit temperature as high as 1600 K. We have also successfully developed a micro turbine device, which is equipped with enhanced micro air-bearings and driven by compressed air. A rotation speed of 15,000 rpm has been demonstrated during lab test. In this paper, we will introduce our research results major in the development of micro combustor and micro turbine test device.

## 1. INTRODUCTION

As part of the effort in developing micro power generation systems, miniaturization of gas turbine engine using MEMS technology is proposed by MIT [1-3]. Tohoku University has also fabricated three-dimensional micro turbines using micro milling [4-5]. In micro heat engine system, both the heat loss and chamber wall cooling in the combustor are critical problems. A design of recirculation gas flow jacket in micro combustor was reported in [3], which had limited effects in preheating the fuel/air mixture and cooling the combustor sidewall.

Our research aims to develop a micro power generation systems based on micro gas turbine engine and a piezoelectric converter, as illustrated in Fig. 1 [6]. The micro gas turbine engine is composed of a centrifugal compressor, a combustor and a radial inflow turbine. The piezoelectric converter is to produce electricity from the rotation of the turbine, which links to the piezoelectric element.

In this paper, we present our research on micro combustor, which consists of seven layers of silicon structures. An extended gas path is applied in micro combustor design, which has been proved to be effective for sustaining high temperature in combustion chamber and for cooling the chamber sidewall. Stable combustion is sustained in the micro chamber, and exit temperature has achieved as high as 1600 K [7]. We have also developed a turbine device with enhanced micro air-bearings, which has achieved a rotation speed of 15,000 rpm. In this paper, we will introduce the concept of micro power generator, the system design of micro gas turbine engine, and some investigations on critical components.

## 2. THE ENGINE SYSTEM DESIGN

We aim to use hydrogen fuel in our micro gas turbine engine. The simplified configuration of the micro power generator is illustrated in Fig 2. The gas turbine engine has a dimension of 21.5 mm × 21.5 mm. It will be made from seven layers of silicon wafers with a total thickness of about 4.4 mm after assembly. Both of the compressor and turbine consist of centrifugal blades with two-dimensional profiles.

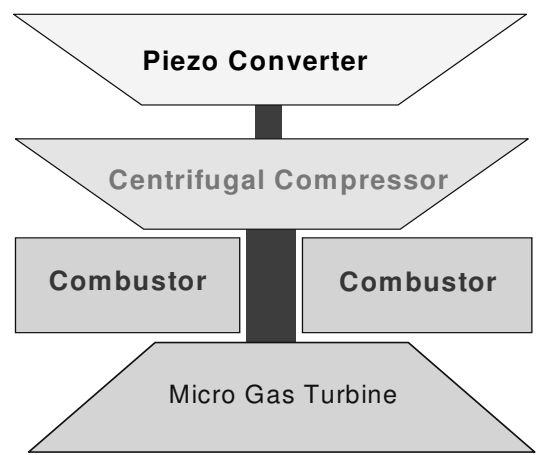

Fig. 1 Schematic of micro power generation system

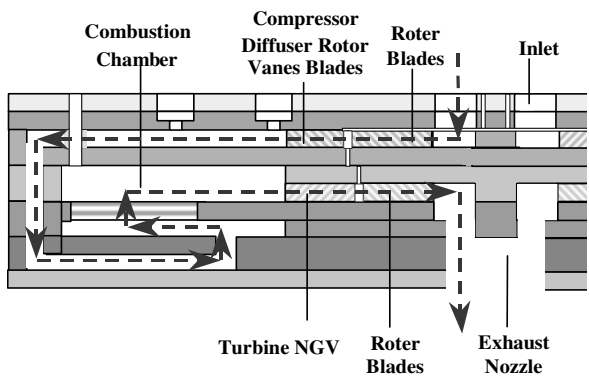

Fig. 2 Implementation of the micro gas turbine engine

In the system design, the inlet airflow rate, compressor ratio and consumption rate of hydrogen fuel are targeted at 0.36 g/s, 4:1 and 17 g/h, respectively. The adiabatic efficiencies of compressor and turbine are estimated to be 0.65 and 0.75, respectively. The efficiency and pressure recovery coefficient of combustor are predicted to be 0.74 and 0.92, respectively [6]. The net power output of the designed gas turbine engine is expected to be 39 watts.

## 3. MICRO COMBUSTOR

### 3.1. Design of micro combustor

The micro combustor [7] utilizes the design of a static gas turbine engine, which means that the combustor has almost the same configuration as the gas turbine engine, as shown in Fig.2, but without the rotating blades. The micro combustor is also based on seven layers of silicon wafers and uses hydrogen as the fuel. The function of each layer is expressed concisely as below.

Fuel is injected through a hole-array on the $2^{nd}$ wafer, and mixes with air and preheated as it flows through recirculation channel made up of the $3^{rd}$ to $6^{th}$ wafers. A hairpin-shaped channel on the $6^{th}$ wafer is designed to prolong the gas flow path in order to (i) preheat the fuel/air mixture well, (ii) efficiently cool down the outer wall of combustor, (iii) reduce heat loss via outer wall and (iv) sustain a stable flame. Then the mixture is injected into the combustion chamber through a set of flame holders on the $5^{th}$ wafer, reacts in the annular combustion chamber, and finally exhausts through the turbine vans on the $4^{th}$ wafer.

Comprehensive CFD (Computational Fluid Dynamics) simulations have been performed to investigate the mixing and combustion characteristics of hydrogen and air in the micro-combustor system. The simulation takes into account the coupling of fluid dynamics, heat transfer and detailed chemical kinetics. The fluid dynamics and heat transfers within the micro combustor, as well as the conjugated heat transfer in solid engine walls, are simulated using Fluent 6.0, and the detailed chemical kinetics of hydrogen-air combustion is expressed by DETCHEM as the user-defined functions of Fluent [8].

Fig. 3 shows some simulation results of temperature distribution inside the combustor, where the flow rate and equivalence ratio were set to be 0.15 g/sec and 0.6, respectively. The heights of combustor chamber were set to 0.6 mm and 1.0 mm, respectively. It was found that, when the equivalence ratio was as high as 0.6, the flame can be stable in the combustor for these cases. The temperature peak of flame and wall temperature and the combustion efficiency are very similar in these two cases. However, when the equivalence ratio was set at 0.5 and chamber height decreased to 0.6 mm, the flame can not be sustained in the combustor. The CFD simulations also show that the hairpin-shaped design of recirculation channel is effective for sustaining higher temperature inside combustion chamber and for cooling the outer walls.

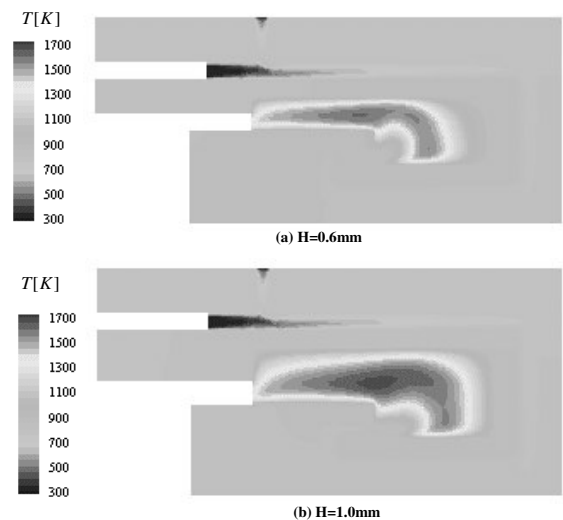

Fig. 3 Temperature distribution inside micro combustor under different chamber heights

### 3.2. Fabrication and characterization

The 7-layer implementation of micro combustor with a hairpin shaped recirculation channel has been successfully fabricated from silicon wafers. DRIE (Deep Reactive Ion Etching) is the major process used in the fabrication of micro combustor. Silicon wafers of 0.4 mm thick are used for the $1^{st}$, $2^{nd}$ and $7^{th}$ layer, while wafers of 0.8 mm thick are used for the $3^{rd}$ to $6^{th}$ layers. Fig. 4 shows the fabricated structures before assembly, while Fig. 5 shows the cross-section of the micro combustor after assembly, which has a size of $21 \times 21 \times 4.4$ mm$^3$. The height of the combustor chamber was set to 0.8 mm at first, which changed to 1.2 mm by adding an extra wafer in order to investigate the size effects of combustor chamber.

(a) View from top-side

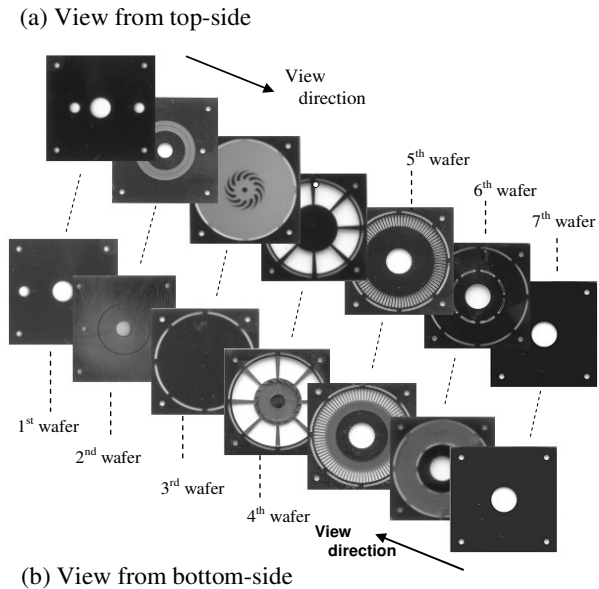

(b) View from bottom-side

Fig. 4 Seven silicon dies of a micro combustor

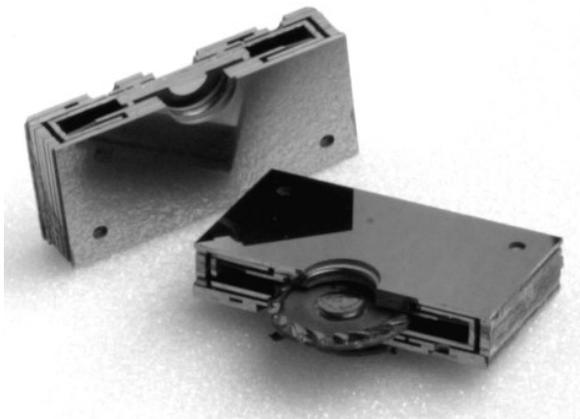

Fig. 5 Cross-section views of micro combustor

To investigate the temperature profile during combustion, the assembled dies are assembled on a stainless steel fixture, which facilitates the connection with fuel/air supply and pressure sensors. Temperature was measured using a 0.5 mm-diameter K-type thermal coupler. Fig. 6 depicts the recorded temperature versus mass flow rate of air/fuel mixture when the equivalence ratio is kept at 0.8. The results illustrate that stable combustion can be sustained when mass flow rate is over 0.04 g/sec. The temperatures near the edge of the exit port range from 1400 to 1600 K.

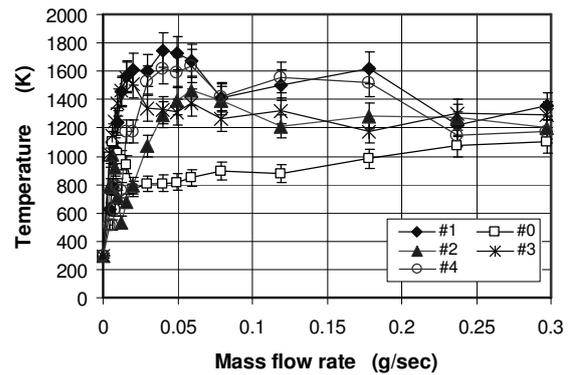

Fig.6 Recorded temperatures at each position versus mass flow rate with a chamber height of 1.2 mm (#1: centre of gas exit; #0, #2-4: near exit edge)

## 4. MICOR TUBINE DEVICE

### 4.1. Design of micro turbine

Micro turbine, a critical part in micro gas turbine engine [9-10], is used for generating power and driving the compressor. In our design, the micro turbine is made up of three layers of silicon structure. Two acrylic plates are used for tubing and clamping. Fig. 7 shows the schematic diagram of our test rig for micro turbine. There are three flow paths driving for rotor, journal air bearing and thrust air bearing, respectively.

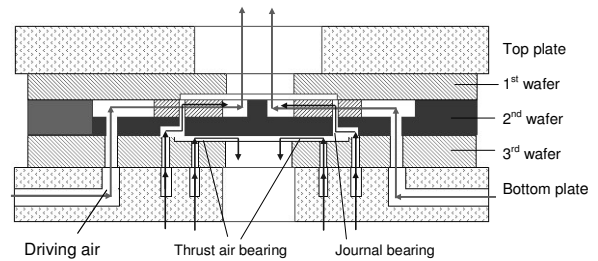

Fig. 7 Design of micro turbine testing device. The turbine device consists of three layers of silicon structure and two acrylic plates.

Silicon wafers of 0.4 mm thick are used for $1^{st}$ and $3^{rd}$ layers of wafers. A wafer of 0.8 mm thick is used for the $2^{nd}$ wafer, from which the rotor and stator are fabricated. The rotor has 17 blades with outer and inner diameters of 8.2 mm and 4.4 mm, respectively. The stator has 23 blades for guide vanes. The blade height is half of the wafer thickness.

The velocity vectors that express the relative flow of driving air at middle section of blades have been investigated by 3-dimensional CFD simulation. It can be seen from Fig. 8 that the airflow velocity in the boundary layer gradually increases from zero to the main flow velocity. The angle of airflow at rotating blade inlet matches the blade profile. This ensures the improvement of efficiency and augmentation of output power. The design shown in Fig. 8 is under the assumption that the turbine is driven by exhausted gas from a combustor. When it is driven by compressed air at room temperature, the rotation speed and output power will be slightly decreased.

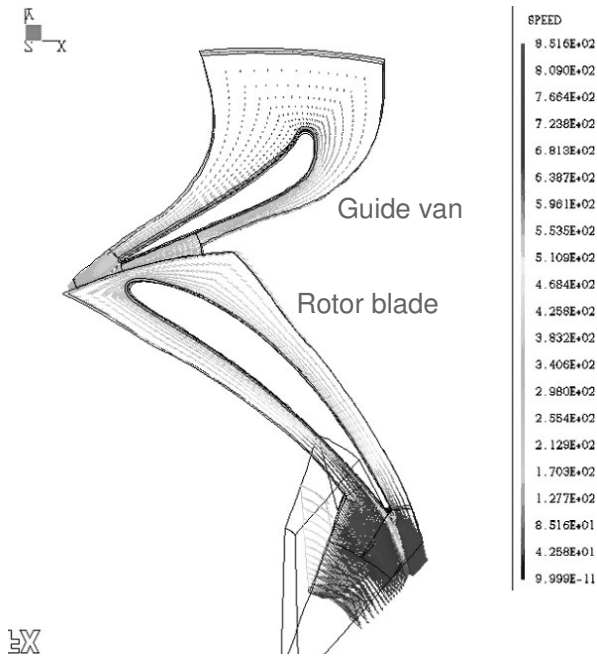

Fig. 8 Turbine blade profiles under CFD simulation

To decrease friction and wear of micro turbine devices during high speed operation, the dynamic and self-acting thrust air bearings are designed and applied in our micro turbine. A pair of thrust air bearings is used for supporting the rotor from its top and bottom. A pump-in configuration with spiral grooves was applied for top thrust air bearing, and a pump-out configuration was applied for bottom bearing. Due to the miniaturized size and small weight of thrust plate, the axial load is mainly introduced by the pressure difference between the top and bottom thrust air bearings.

### 4.2. Fabrication of micro turbine

Similar to micro combustor, DRIE is also the major process for fabrication of micro turbine. However, the non-uniformity of DRIE process in term of etching rate over a 4-inch wafer can be as high as 5% under normal process condition. When etching the 2$^{nd}$ wafer, such non-uniformity may introduce the mismatch of mass center and geometry center of rotors, which will cause collision between rotor and side wall.

To enhance the etching uniformity, we used pre-bonded silicon wafer, which was embedded with a 1.5 µm-thick $SiO_2$ layer. This oxide layer can serve as the etching stopper during DRIE process. The topographical map shown in Fig. 9 illustrates rotor profiles fabricated from a pre-bonded silicon wafer.

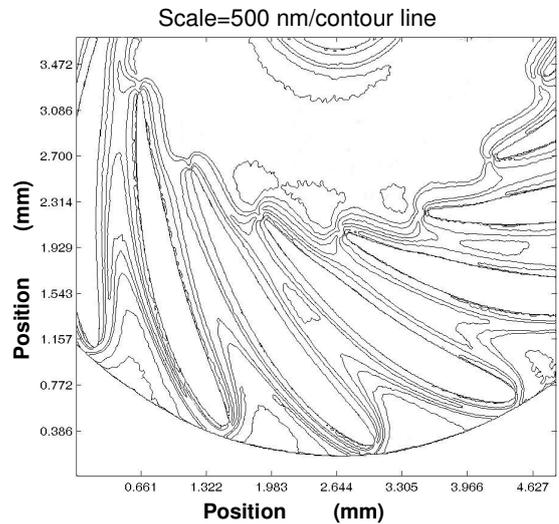

Fig. 9 Topographical map of the rotor blades fabricated from a pre-bonded wafer

A novel structure of thrust air-bearing was generated together with the micro turbine. The top bearing uses pump-in type grooves and was etched out from the bottom side of the 1$^{st}$ wafer. The bottom bearing utilizes pump-out type grooves and is etched out from the top side of the 3$^{rd}$ wafer. The DRIE process for fabricating thrust bearings has been specifically optimized. The etching depths of the grooves on the top and bottom bearings are 15 µm and 36 µm, respectively. Fig. 10 illustrates the fabricated pump-in spiral grooves for top thrust bearing.

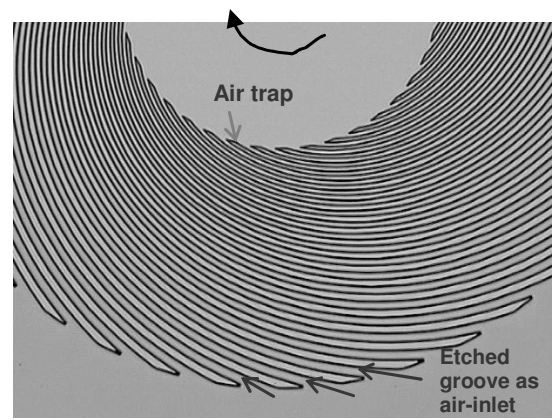

Fig. 10 Pump-in spirals for top dynamic thrust bearing with a etch depth of 15 µm.

### 4.3. Testing of the turbine device

The micro turbine was mounted on an acrylic test jig and connected to a gas distribution system for air-driven rotation testing. Fig. 11 shows the experimental setup. An optical fiber sensor is used to measure the rotation speed of the rotor blade, by sensing the periodic change of reflectivity. A stable rotation of 15,000 rpm is recorded during the experiments.

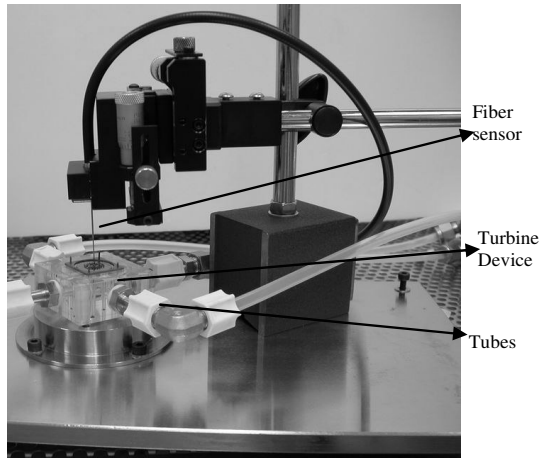

Fig. 11 Lab setup for rotation measurements of micro air-driven turbine

### 5. SUMMARY

Aiming towards a silicon-based micro gas turbine engine for power generation, we have investigated the system design, fabricated the micro combustor and micro turbine. The micro combustor consists of seven layers of silicon structures and uses a hairpin-shaped fuel/air recirculation channel for the extension of flow path. The experimental investigations demonstrate that the combustor can sustain a stable combustion with an exit temperature as high as 1600 K. We applied new structure of micro air-bearings in our micro turbine. A stable rotation with 15,000 rpm has been reported. Our future research will include the integration of turbine and compressor with micro combustor so as to implement the micro gas turbine engine.

### ACKNOWLEDGEMENTS

The authors would like to thank the Agency for Science, Technology and Research (A*Star), Singapore for financial support under the project No. 022 107 0011.